\documentclass{article}
\usepackage{graphicx}
\usepackage{latexsym}
\usepackage{amssymb,amsfonts,amsmath}
\usepackage{epsf}
\usepackage[dvips]{color}
\title{The decay of unstable k-strings in SU(N) gauge theories at zero and 
finite temperature}
\author{Ferdinando Gliozzi}
\newcommand{\eq}{\begin{equation}}
\newcommand{\en}{\end{equation}}
\newcommand{\ear}{\begin{eqnarray}}
\newcommand{\rae}{\end{eqnarray}}
\newcommand{\Z}{\mathbb{Z}}
\newcommand{\C}{\mathbb{C}}
\newcommand{\uu}{\mathbb{I}}
\newcommand{\R}{{\cal R}}
\newcommand{\s}{{\cal S}}
\newcommand{\bra}{\langle}
\newcommand{\ket}{\rangle}
\newcommand{\um}{\frac12}
\newcommand{\tr}{{\rm tr}\,}
\newcommand{\Tr}{{\rm tr}^{~}}
\definecolor{M_Beige}         {rgb}{0.96 , 0.96 , 0.86}

\definecolor{M_Brown}         {rgb}{0.65 , 0.16 , 0.16}

\definecolor{M_Gold}          {rgb}{1.00 , 0.84 , 0.00}

\definecolor{M_LemonChiffon}  {rgb}{1.00 , 0.98 , 0.80}

\definecolor{M_Orange}        {rgb}{1.00 , 0.60 , 0.00}

\definecolor{M_Pink}          {rgb}{1.00 , 0.75 , 0.80}

\definecolor{M_Violet}        {rgb}{0.93 , 0.51 , 0.93}

\newcommand{\Cblu}[1]{\textcolor{blue}{#1}}

\begin{document}
\maketitle
\noindent
Dipartimento di Fisica Teorica, Universit\`a di Torino and\\ INFN,
sezione di Torino, via P. Giuria, 1, I-10125 Torino, Italy.\\
\vskip0.5cm\noindent
e-mail: gliozzi@to.infn.it\\
\begin{abstract}
Sources in higher representations of $SU(N)$ gauge theory at $T=0$ 
couple with apparently stable strings with tensions depending on 
the specific representation rather than on its $N-$ality. Similarly
at the deconfining temperature these sources carry their own 
representation-dependent critical exponents. It is pointed out that in 
some instances one can evaluate exactly these exponents by fully 
exploiting the correspondence between the 2+1 dimensional critical 
gauge theory and the $2d$ conformal field theory in the same universality 
class. The emerging functional form of the Polyakov-line correlators 
suggests a similar form 
for Wilson loops in higher representations which helps in understanding 
the behaviour of unstable strings at $T=0$. A generalised Wilson loop in 
which along part of its trajectory a source is converted in a gauge 
invariant way into higher representations with  same $N-$ality could 
be used as a tool to estimate the decay scale of the unstable strings.   

\end{abstract}

\section{Introduction}

One of the most fascinating aspects of numerical experiments in lattice 
gauge theory is the possibility to make many controlled changes to 
explore the response of the strong interaction dynamics. In particular, we 
can vary the quark masses, the number $N$ of colours and even remove
the sea quarks. In this way one is led to study pure $SU(N)$ gauge theory
to probe the main properties of the confining vacuum.   

In this context, the linear rising of the static potential between a pair 
of quarks in the fundamental representation is well described  
by a thin flux tube, or string, joining the two quarks.
Excited colour sources, {\it i.e.} sources in a higher representation $\R$ of 
the gauge group, behave at intermediate distances in a similar way, 
giving rise to the formation of a confining string with a string tension 
$\sigma_\R$. However most strings of this kind are expected to be unstable: the
long distance properties of the string attached  to a source in a 
representation $\R$ built up of $j$ copies of the fundamental representation 
should depend only on its $N-$ality $k^{~}_\R\equiv j~({\rm mod} N)$ because 
all representations with same $k$ can be converted into each other by the 
emission of a proper number of soft gluons. As a consequence, the heavier 
$\R-$strings are expected to decay into the string with smallest string 
tension within the same $N$-ality class, called $k-$string. 
For a recent discussion on this subject see \cite{as}.

General heuristic 
arguments suggest that stable $k-$strings belong to the anti-symmetric 
representation with $k$ quarks, as it has been supported by exact results 
in various approaches to ${\cal N}=1$ supersymmetric  $SU(N)$ gauge theories 
\cite{ds,hsz,hk} as well as by  lattice calculations in some $SU(N)$ pure 
gauge models \cite{dprv,lt,ltw,dprv2}.  

In this paper we address the question of stability of strings 
attached to sources in different representations with the same $N-$ality.
Although most numerical experiments based on large Wilson loops 
\cite{cj,pt,sd,ba,cp} seem to defy the above theoretical arguments, 
yielding  apparently stable string tensions which depend on $\R$ rather 
than on its $N-$ality, this kind of behaviour has been fully understood, 
at least in the case of the adjoint string. Since the $N-$ality of the adjoint 
representation is zero, the static potential of two adjoint sources is 
expected to level off at large 
separations. Correspondingly, the associated string should decay into 
a pair of bound sates of a static adjoint colour source and a gluon field, 
sometimes called glue-lump in the literature. Knowing the mass of the glue-lumps
one can even evaluate the scale $r_{adj}$ at which the adjoint string 
breaks \cite{pt}. 

The lack of any sign of adjoint string breaking 
in the above-mentioned studies, while measuring the static potential from  
Wilson loops only, indicates that such an  operator has  a  
poor overlap with the true ground state. This fact has been directly 
demonstrated  in $2+1$ $SU(2)$ gauge theory  \cite{krde} where,
using a variance reduction algorithm allowing to detect signals down to 
$10^{-40}$, it has been clearly observed a rectangular Wilson loop 
 $W(r>r_{adj},t)$
 changing sharply its slope as a function of $t$ from that associated to 
the unbroken string (area-law decay) to that of the broken-string state 
(perimeter-law decay) at a distance much longer than the adjoint string 
breaking scale $r_{adj}$. 

\begin{figure}
\centering
\includegraphics[width=0.9\textwidth]{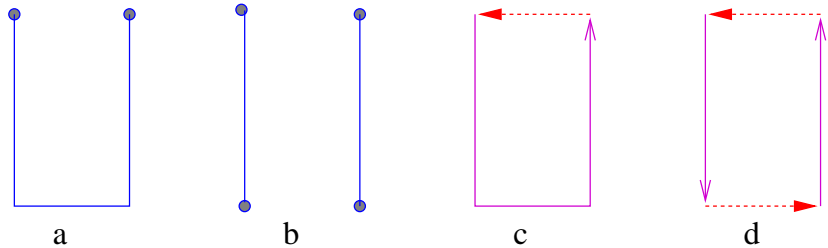} 
\caption{Glue-lump operators and mixed Wilson lines. The operators depicted in 
$a$ and $b$ describe adjoint Wilson lines decaying in the vacuum. They 
are used to extract the adjoint static potential in the multichannel method. 
The operators $c$ and $d$, described in the text, generalise the glue-lump operators to the case of representations of non zero $N-$ality .} 
\label{Figure:1}
\end{figure}

Alternatively, one can enlarge the basis of the operators used to extract 
the adjoint potential 
in order to find a better overlap to the true ground state, following a 
multichannel method originally advocated in \cite{cm}. 
Indeed adjoint sources, contrarily to what happens in the case of fundamental 
representation, can form gauge invariant \Cblu{open} Wilson lines, like those 
depicted in Fig.\ref{Figure:1} $a$ and $b$, having a good overlap with 
the two glue-lump state. In this way a rather abrupt crossover between 
string-like and broken string states has been clearly seen at the expected 
distance $r_{adj}$ both in $2+1$ \cite{st} and $3+1$ \cite{fp} $SU(2)$ 
gauge models.
The same multichannel method allowed to observe breaking of
fundamental string in various gauge theories coupled to dynamical matter
fields in 2+1 \cite{pw} and 3+1 dimensions \cite{ks}, including QCD with two 
flavours \cite{bdl}. Also in these cases ordinary Wilson loops do not show 
clear signs of string breaking even at sizes much larger than the expected
breaking scale, except in a particularly simple case, the 2+1 
$\Z_2$ gauge-Higgs model \cite{Gliozzi:2004cs}, where, using  a 
variance reduction algorithm like in \cite{krde}, fundamental string 
breaking has been convincingly demonstrated.     
\par
One of the results of the present work is the generalisation of the multichannel method to the decay of unstable strings of non-vanishing $N-$ality.
To this aim we define in Sect.\ref{mixed} a new kind of gauge-invariant 
operators, the \Cblu{mixed Wilson loops}, (see Fig.\ref{Figure:1} $c$ and $d$)
where  along one or more segments of the closed path $\gamma$ 
(the dashed lines in Fig.\ref{Figure:1}) the static source carries the 
quantum numbers of an excited representation $\R$ of $N-$ality $k$, 
while in the remaining path 
(solid lines) the source lies in the stable, fully anti-symmetric 
representation. These operators have a good overlap with the stable 
$k-$string state and constitute the most obvious generalisation of the 
glue-lump operators drawn in Fig.\ref{Figure:1} $a$ and $b$. 
\par The above construction follows a general discussion  of the expected 
overlap properties of ordinary Wilson loops in excited representation. 
This leads to a rough estimate of the decay scale of the unstable 
strings and offers a simple explanation of the apparent stability of 
these strings at $T=0$.
\vskip .3 cm   
\par
The $T>0$ situation is less problematic. According to  a simple diagrammatic 
argument \cite{gp}
the correlator of  Polyakov lines is expected to have a good 
overlap with the true ground state. These correlators  
 in non-fundamental representations have been studied in 
various instances \cite{cb,ot,Muller:1991xj,DelDebbio:2004zc}
and it has easily observed the screening of the adjoint representation, 
and the decay of excited representations of non vanishing triality in $SU(3)$
and even the decay of the symmetric 2-index representation into the 
anti-symmetric one in $SU(4)$ \cite{DelDebbio:2003tk}. A difficulty 
of the $T>0$ approach is to extract reliable estimates of the breaking 
or decay scales of unstable strings at $T=0$.
\vskip .3 cm
\par
A much more challenging problem emerges at the deconfining point: in gauge 
theories with continuous  phase transition one 
obvious question concerns the critical behaviour of the Polyakov lines 
in arbitrary representations. Over the years,many studies have been 
dedicated to this subject  \cite{gw,dam,jk,ctd,dh,pi}. 
\par The surprising result is 
that sources of higher representations, according to various numerical 
experiments \cite{dam,ctd,dh}, correspond to \Cblu{different} magnetisation 
exponents, one exponent for each representation. Universality arguments would,
for such continuous phase transitions, place the finite-temperature $SU(N)$ 
gauge theory in the universality class of $\Z_N$ invariant spin model in one 
dimension less \cite{sy}. 
The spin operator is mapped into the Polyakov line in the fundamental
representation. What about the Polyakov lines in higher representations of 
$SU(N)$? There appears to be no room for independent exponents for these 
higher representations from the point of view of the abelian spin system, 
since there is simply no obvious analogue of the non-abelian process of 
soft gluons emission, creating higher representations in the same $N-$ality 
class. 

Actually a mean field approximation of the the effective $SU(2)$ 
Polyakov-line action at criticality in the $d\to \infty$ limit shows that the
leading amplitudes of higher representations vanish at strong coupling, 
and the sub-leading exponents become dominant, thus each higher representation 
source carries its own critical exponent\cite{dh}.      

Here we reach a similar conclusion for $SU(3)$ and $SU(4)$  critical 
theories in $2+1$ dimensions starting from a different point of view:
we find a map between the operator product expansion (OPE) of the Polyakov 
operators in the gauge theory and the corresponding spin operators in the 
two-dimensional conformal field theory which describes the associated 
spin system at criticality. As a result, we are able to conjecture, 
for instance, an exact value for the $\eta$ exponent (or anomalous dimension) 
of the sources in the \Cblu{\sl symmetric}  representation made with 
two quarks (Sect.\ref{polyakov}). The resulting functional form of the 
Polyakov-Polyakov correlator at criticality of this excited representation 
is the starting point of an Ansatz for the vacuum expectation value 
of the  Wilson loop associated to sources in higher representations at $T=0$
(Sect.\ref{decay}). Section \ref{mixed} is dedicated to 
the construction of mixed Wilson loops and finally in Sect.
\ref{conclusion} we draw some conclusions.       
\section{Polyakov loops at criticality}
\label{polyakov}
Consider a $d+1$ dimensional pure gauge theory undergoing a continuous 
deconfinement transition at the critical temperature $T_c$. The effective model describing the behaviour of Polyakov lines at finite $T$ will be  a
 $d-$dimensional spin model with a global symmetry group coinciding with 
the center of the gauge group. Svetitsky and Yaffe \cite{sy} (SY) were 
able to show that this effective model has only short-range 
interactions, then it follows from universality arguments that the 
spin model belongs in the same universality class of the original 
gauge model and the high temperature phase of the gauge theory is like the 
low temperature phase of the spin system.

It is clear that this SY conjecture, which has passed several numerical 
tests, becomes very predictive for $d=2$, where, using the methods of 
conformal field theory (CFT), the critical behaviour can be determined 
exactly. For example, the critical properties of $2+1$ dimensional $SU(3)$ 
gauge theory at deconfinement coincide with those of the 3-state Potts model,
as it has been checked in numerical simulations \cite{ctd}.

What is needed to fully exploit the predictive power of the SY conjecture 
is a mapping relating the physical observables of the gauge theory to the 
operators of the dimensionally reduced model, as first advocated in
\cite{Gliozzi:1997yc}.

The correspondence between the  Polyakov line 
in the fundamental representation
$f$ and the order parameter $\sigma$ of the spin model is the first entry  
in this mapping:
\eq
\Tr_f(U_{\vec{x}}) ~\sim~ \sigma(\vec{x})~, 
\en
$U_{\vec{x}}$ is the gauge group element  associated to the closed path 
winding once around the periodic imaginary time direction intersecting 
the spatial plane at the point $\vec{x}$. The above equivalence is only 
valid in a weak sense, that is, when the left-hand-side of the equation  
is inserted in a correlation function of the gauge theory and the right-hand 
side in the corresponding correlator of the spin model.
 \par It is now natural to ask what operators in the CFT correspond to
Polyakov lines in  higher representations. \Cblu{On the gauge side} 
these can be obtained by a proper combination of products of Polyakov 
lines in the fundamental representation, using repeatedly  the  general 
property
\eq
\Tr_\R(U)\,\Tr_\s(U)=\sum_{\R'\in\R\otimes\s}\Tr_{\R'}(U)~,
\label{decompo}
\en  
valid for any pair of irreducible representations of an arbitrary group.
In order to control the  singularities in the correlator functions 
due to evaluation of local operators at the same point 
we may resort to the operator product expansion (OPE). The OPE of Polyakov operator in the fundamental representation can be written in the form
\eq
  \Tr_f(U_{\vec{x}})\; \tr_f(U_{\vec{y}})=
\sum_{\R\in f\otimes f}C_\R(\vert\vec{x}-\vec{y}\vert)\,\Tr_{\R}(U_{(\vec{x}+\vec{y})/2})
+\dots
\en
where the coefficients $C_\R(r)$ are suitable functions (they become  powers
of $r$ at the critical point) and the dots represent the contribution of 
higher dimensional local operators. The important property of  this OPE
is that the local operators are classified according to the irreducible 
representations of $G$ obtained by the decomposition of the direct product 
of the representations of the two local operators in the left-hand side.

\Cblu{On the CFT side} we have a similar structure.  The order parameter
$\sigma$ belongs to an irreducible representation $[\sigma]$ of the Virasoro 
algebra \cite{bpz} and the local operators contributing to an OPE are 
classified according to the decomposition of the direct product of the 
Virasoro representations of the left-hand-side operators. This decomposition 
is known as fusion algebra and can be written generically as
\eq
[\chi_i]\star[\chi_j]= c_{ij}^k[\chi_k]
\label{fusion}
\en
where the integers $ c_{ij}^k$  are the fusion coefficients.
In the case of three-state Potts model there is  a finite number of 
representations that we list along with their scaling dimensions
\footnote{Actually the critical three-state Potts model is invariant under
a larger algebra than that of Virasoro, the so-called ${\cal W}_3$ algebra,
and the representations listed in (\ref{potts}) are irreducible 
representations of such a larger algebra. A complete list of 
these fusion rules can be found for instance in \cite{Fuchs:1998qn}.}

\eq
\begin{matrix}
{\uu}~({\rm identity});& \sigma,\,\sigma^+~({\rm spin~fields});&
\epsilon~({\rm energy});&\psi,\psi^+\\
~&~&~&\\
x_{\uu}=0&x_\sigma=\frac2{15}& x_\epsilon=
\frac4{15}& x_\psi=\frac43\\
\end{matrix}
\label{potts}
\en 

The fusion rules we need are
\eq
[\sigma]\star[\sigma]=[\sigma^+]+[\psi^+]~;~\,[\sigma]\star[\sigma^+]=
[{\uu}]+[\epsilon]~;~\,[\psi]\star[\epsilon]=[\sigma]~.
\label{fusion3}
\en 

Comparison of the first equation with the analogous one of the gauge side
\eq
\{3\}\otimes\{3\}=\{\bar{3}\}+\{6\}~,
\en
owing to the correspondence $\Tr_{\bar{f}}(U_{\vec{x}})\sim 
\sigma^+(\vec{x})$, yields a new entry of the gauge/CFT correspondence
 \eq
\Tr_{\{6\}}(U_{\vec{x}}) ~\sim~ \psi^+(\vec{x})+
c\,\sigma^+(\vec{x})~,
\label{c6}
\en
where there are no a priori reasons for the vanishing of the 
coefficient $c$. Hence the Polyakov-Polyakov critical correlator
of the symmetric representation $\{6\}$ is expected to have the 
following general form in the thermodynamic limit
\eq
\bra\Tr_{\{6\}}(U_{\vec{x}})\;\Tr_{\{\bar{6}\}}(U_{\vec{y}})
\ket_{SU(3)}^{~}=
\frac{c_s}{r^{2x_\sigma}}+\frac{c_u}{r^{2x_\psi}}~,
\label{c66}
\en
 with $r=\vert\vec{x}-\vec{y}\vert$ and $c_s,c_u$ suitable coefficients.
Since $x_\sigma<x_\psi$, the second term drops off more rapidly than the 
first, thus at large distance this correlator behaves like that of 
the anti-symmetric representation $\{\bar{3}\}$ as expected also at zero 
temperature. 

A similar reasoning  can be applied to the second  fusion rule 
(\ref{fusion3}), which is the general form relating the order parameter to 
the energy operator in spin systems. The gauge side of this equation is
\eq
\{3\}\otimes\{\bar{3}\}=\{1\}+\{8\}
\label{b33}
\en
which leads to the new entry
\eq
\Tr_{adj}(U_{\vec{x}})~\sim~ a+\epsilon(\vec{x})~,
\label{cadj}
\en
which is expected to be valid for any $SU(N)$ gauge theory undergoing 
a continuous phase transition.
The constant $a$ can be numerically evaluated\cite{jk} using the 
expected finite-size behaviour  
\eq
\bra\Tr_{adj}(U)\ket=a+\frac b{L^{2-1/\nu}}~, 
\en
where $L$ is the spatial size of the system and we used the general relation 
$x_\epsilon=d-1/\nu$ relating the scaling 
dimension of the energy operator to the thermal exponent  $\nu$. 

Finally the third fusion rule ( \ref{fusion3}), combined with (\ref{cadj}), 
can be interpreted as the CFT counterpart of the soft gluon emission which 
converts into each other stable $(\sim\sigma)$ and unstable $(\sim\psi)$
strings.
\vskip .3 cm

Another example of this gauge/CFT correspondence can be worked out in the 
2+1 dimensional $SU(4)$ gauge model, where now the $2d$ CFT is the symmetric
Ashkin-Teller model \cite{Kadanoff:1978pv}. It describes two  Ising 
models with spin fields $\sigma(\vec{x})$ and $\tau(\vec{x})$ with scaling 
dimensions $x_\sigma=x_\tau=\frac18$ coupled through a local four spin 
interaction which depends on a coupling constant $g$. 

The model is invariant under  three different $\Z_2$ transformations:
\eq
\sigma\to-\sigma~;~~\tau\to-\tau~;~~\sigma\leftrightarrow\tau~.
\label{sym}
\en
The first two correspond to symmetries that can be spontaneously broken 
by  a thermal variation, leading to an order-disorder transition. Hence 
they should be associated to the center $\Z_4$ of the gauge group. The  
$\sigma\leftrightarrow\tau $ symmetry, on the contrary, cannot 
be spontaneously broken by  a thermal variation  and corresponds 
to charge conjugation. 
The first entry of the gauge/CFT correspondence can therefore be written
in the form
\eq
\Tr_{\{4\}}(U_{\vec{x}}) ~\sim~ \theta(\vec{x})=
e^{-i\pi/4}\,\left(\sigma(\vec{x})+i\tau(\vec{x})\right)~. 
\label{f4}
\en    
The phase $\pi/4$ comes from the observation that  
charge conjugation $ \sigma\leftrightarrow\tau $ should correspond, on the 
gauge side, to  complex conjugation; this gives the constraint
\eq
\begin{matrix} \theta(\vec{x})& \longrightarrow &\theta^*(\vec{x})\\
~& \sigma\leftrightarrow\tau &~\\
\end{matrix}~,
\en
which fixes such a phase factor.

The local operators which occur in the OPE of 
$\theta(\vec{x})\;\theta(\vec{y})$ or 
$\theta(\vec{x})\;\theta^*(\vec{y})$
are the so-called ``ploarization'' $\pi(\vec{x})\in[\pi]=[\sigma]\star[\tau]$
and the ``cross-over'' operator   $\rho(\vec{x})\in[\rho]=
[\sigma]\star[\sigma]-[\tau]*[\tau]$ as well as the energy $\epsilon$ 
and the identity.

The Ashkin-Teller model is not an isolated 
critical system like the three-state Potts model, but describes a 
line of fixed points depending on the coupling parameter $g$.
While the spin fields retain along the fixed line their scaling dimensions,
$\epsilon$, $\pi$ and $\rho$ have $g-$dependent scaling 
dimensions. They obey, however, the simple relations
\eq
\frac{x_\pi}{x_\epsilon}=\frac14~,~~x_\rho\,x_\epsilon=1~.
\en
Numerical experiments \cite{deForcrand:2003wa} indicate that the critical 
$SU(4)$ gauge theory is located near the four-state Potts model, 
corresponding to $x_\epsilon=\um$  \cite{Kadanoff:1978pv}.

Let us consider the $SU(4)$ representations made with two quarks. We have
\eq
\{4\}\otimes\{4\}=\{6\}+\{10\}~,
\en
where the anti-symmetric representation  $\{6\}$ is real, 
$i.e.$ $[\Tr_{\{6\}}(U)]^*=  \Tr_{\{6\}}(U)$, being the vector 
representation of $SO(6)\sim SU(4)$. On the CFT side, using (\ref{f4})
in the OPE of   $\theta(\vec{x})\;\theta(\vec{y})$   yields at once
\eq
[\theta]\star[\theta]=[\pi]+[\rho]~,
\en 
where $[\pi]=[\sigma]\star[\tau]$  is real and even under charge 
conjugation  $\sigma\leftrightarrow\tau$, while $[\rho]$ is purely imaginary and odd. We have therefore 
two new entries for the $SU(4)$/CFT correspondence
\eq
\begin{split}
\Tr_{\{6\}}(U_{\vec{x}})~\sim ~\pi(\vec{x})~,\\
\Tr_{\{10\}}(U_{\vec{x}})~\sim ~\rho(\vec{x})+c\,\pi(\vec{x})~,
\end{split}
\en 
where $c$ must be different from zero, because $\Tr_{\{10\}}(U)$ 
is not purely imaginary.

In analogy with Eq.(\ref{c66}) we can write
\eq
\bra\Tr_{\{10\}}(U_{\vec{x}})\,\,\Tr_{\{\bar{10}\}}(U_{\vec{y}})
\ket^{~}_{SU(4)}=
\frac{c_s}{r^{x_\epsilon/2}}+\frac{c_u}{r^{2/x_\epsilon} }~,
\label{c10}
\en
with both $c_s$ and $c_u$ are different from zero. This
 shows that  the Polyakov-Polyakov correlator  in the symmetric 
representation with $N-$ality 2, even if at short distance is controlled 
by the irrelevant operator $\rho$, at large 
distance  behaves exactly like the  anti-symmetric rep. with the 
same $N-$ality, as expected under physical grounds.
\section{Decay of unstable strings at zero temperature} 
\label{decay}

 As mentioned in the introduction, the difficulty in observing  string 
breaking  or string decay with the Wilson loop seems to indicate nothing 
more than that it has a very small overlap with the broken-string or 
stable string state.
Why? being a general phenomenon which occurs for any gauge group, 
including $\Z_2$, in pure gauge models as well as in models coupled 
to whatever kind of matter, it requires a general explanation which should 
not depend on detailed dynamical properties of the model. 
A simple explanation in the case of gauge models coupled with matter was 
proposed in  \cite{Gliozzi:2004cs}, which now we enforce in the 
present case.

The general form of the Polyakov correlator in higher representations of 
$SU(N)$ found in Eq.s(\ref{c66}) and (\ref{c10}) suggests a simple  
Ansatz describing the asymptotic functional form
of the vacuum expectation value of a large, rectangular, Wilson loop in 
a higher representation $\R$ coupled to an unstable string which should 
decay into a stable  $k-$string
\footnote{For sake of simplicity we neglect the $1/r$ term in the potential 
which accounts for the quantum fluctuations of the flux tube.}
\eq 
\bra W_\R(r,t)\ket\simeq c_u\exp[-2\mu_{\R}^{~}(r+t)-\sigma_{\R}^{~} rt]+c_s
\exp[-2\mu_\R' (r+t)-\sigma_k^{~} rt]
\label{Ansatz} 
\en
Similar proposals are described in \cite{krde,as}.
The first term describes the typical area-law decay produced at intermediate 
distances by the unstable string with tension $\sigma_{\R}^{~}$. 
The second term is  instead the contribution expected 
by the stable $k-$string in which the $\R-$string decays. In the case 
of adjoint representation (zero $N-$ality) 
one has $\sigma_0=0$ and the perimeter term $\mu'_{adj}$ denotes
the mass of the lowest glue-lump. Eq.(\ref{Ansatz}) has to be understood as an 
asymptotic expansion  which approximates  $\bra W_\R(r,t)\ket$ when
$r,t>r_o$, where $r_o$ may be interpreted as the  scale where the confining 
string forms.
\begin{figure}
\centering
\includegraphics[width=0.9\textwidth]{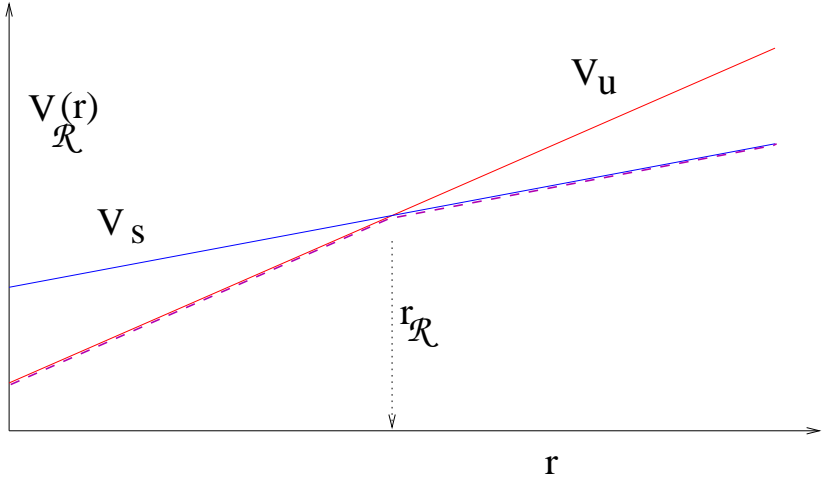} 
\caption{A schematic view of the static potential between sources belonging to 
an excited representation $\R$ (dashed line). 
$V_u$ is the potential experienced at 
intermediate distances generated by the unstable string. $V_s$ is the 
asymptotic behaviour, controlled by the string tension $\sigma^{~}_k$ 
of the stable string in which the unstable string decays. $r^{~}_\R$ is 
the decay scale.} 
\label{Figure:2}
\end{figure}

When $t$ and $r$ are sufficiently large, no 
matter how small $c_s$ is, the above Ansatz 
implies that at  long distances the stable string  
eventually prevails, since 
$\Delta\sigma\equiv\sigma_\R-\sigma_k>0$, hence the first
 term drops off more rapidly than the second. With the emergence of this 
long-distance effect a closely related question comes in: at what distance 
the stable string shows up? it depends on the  
difference $\Delta\mu\equiv\mu'_\R-\mu^{~}_\R$. Since
\eq
-\frac1t\log\bra W_\R(r,t)\ket=2\mu'_\R+\sigma_k^{~}r-
\frac1t\log\left[c_s+{c_u}e^{r2\Delta\mu}
e^{-t(r\Delta\sigma -2\Delta\mu)}\right]~,
\label{logw}
\en
we have  (see Fig.\ref{Figure:2})
\eq
V_\R(r)=V_s(r)=2\mu'_\R+\sigma_k^{~}r~,~~~ r>r_\R^{~}~,
\label{pots}
\en
where 
\eq
V_\R(r)=-\lim_{t \to\infty}
\frac1t\log\bra W_\R(r,t)\ket
\label{poto} 
\en
 is the static potential 
and $r_\R^{~}$ is the  value of $r$ which annihilates the exponent in
Eq.(\ref{logw}) 
\eq
r_\R^{~}=\frac{2\Delta \mu}{\Delta \sigma}\equiv2\frac{\mu'_\R-\mu^{~}_\R}
{\sigma^{~}_\R-\sigma^{~}_k}~.
\label{scale}
\en
Creation of unstable strings requires  $\mu'_\R>\mu^{~}_\R$
(see Fig.\ref{Figure:2}). In the case of 
zero $N-$ality the above equation yields the usual estimate of the adjoint 
string breaking scale. The inclusion of the $\frac1r$ terms in this analysis 
does not modify substantially the numerical estimates.  
 Notice that the mass $\mu_\R^{~}$ and 
$\mu'_\R$ are not UV finite because of the additive  self-energy divergences
(linear in 3+1 dimensions, logarithmic in 2+1 dimensions), which cannot be 
absorbed in a parameter of the theory. However, these divergences should 
cancel in their difference, hence $r^{~}_\R$ is a purely dynamical scale, 
defined  for any non fully antisymmetric representation of $SU(N)$, which 
cannot be tuned by any bare parameter of the theory. 

When $r$ is less than the decay scale $r^{~}_\R$ Eq.(\ref{pots}) is no longer 
valid and is replaced by
\eq
V_\R(r)=V_u(r)=2\mu^{~}_\R+\sigma_\R^{~}r~,~~~ r_o<r<r_\R^{~}~.
\label{potu}
\en
Thus the Ansatz (\ref{Ansatz}) describes the unstable string decay as a level
crossing phenomenon, as observed in the adjoint string.
\par
It is worth noting that, though $c_u$ and $c_s$ must be non-vanishing 
quantities, they  do not contribute to $V_\R(r)$, whereas  play  a 
fundamental role in the 
possibility to observe string decay when both $r$ \Cblu{and} $t$ are finite. 
It is easy to see that  for  $r<r^{~}_\R$ and $t$ large enough the unstable 
string cannot decay, the reason being that $V_u(r)<V_s(r)$ in this range.
In order to avoid the unphysical behaviour in which the decay is visible 
only in a  finite interval of  $t$,  it is  obvious that $c_s$ cannot be 
too big. More precisely, we must assume
\eq 
c_u\, e^{-2\mu^{~}_\R(r+t)-\sigma_\R^{~} rt}\geq c_s\, 
e^{-2\mu'_\R(r+t)-\sigma_k^{~} rt}~,~  
r_o\leq r\leq r^{~}_\R~,~
t\ge r_o~.
\label{Ineq}
\en
With the help of Eq.(\ref{scale}), this inequality can  be recast into the form
\eq
\log\frac{c_s}{c_u}\leq\Delta\sigma\,
\left[r^{~}_\R(r+t)-rt\right]~~,
~r_o\leq r\leq r^{~}_\R,~t\ge r_o~.
\label{ineq}
\en
To minimise the right-hand-side we put $r=t=r_o$ and get

\eq
\log\frac{c_s}{c_u}\leq\Delta\sigma\,
r_o\,(2r^{~}_\R-r_o)~~.
\label{bound}
\en
Such an upper bound constitutes the main obstruction to observe string decay
in the ordinary Wilson loop $W_\R$. Indeed, when $r>r^{~}_\R$ the 
distance $t^{~}_\R$
where the second term of the Ansatz (\ref{Ansatz}) equals the first, then 
it makes possible to see the decay, is given by
\eq
t^{~}_\R\,(r-r^{~}_\R)=r\,r^{~}_\R-\frac1{\Delta\sigma}
\log\frac{c_s}{c_u}\ge\, r^{~}_\R \,(r- 2r_o)+r_o^2~~,
\en
 which shows that $ t^{~}_\R\gg r^{~}_\R$  unless   $ r\gg r^{~}_\R$,
and we shall argue shortly that $r^{~}_\R> r_{adj}$.
From a 
computational point of view it is very challenging  to reach such 
length scales \footnote {In 3+1 dimensional $SU(2)$ 
gauge model, for instance, $r_{adj}\sim 1.25$ fm \cite{fp}.}
in the measure of $ \bra W_\R(r,t)\ket $. 
This explains why unstable string decay has not yet been observed at zero 
temperature. 
\par The evaluation of $r^{~}_\R$
is  problematic, because  Eq.(\ref{scale})   implies an estimate of
the self-energy $\mu'_\R$, which contributes to the sub-dominant term of the 
Ansatz (\ref{Ansatz}), hence from a numerical point of view it is almost 
hopeless. We shall see that the mixed Wilson loop can be used  as a 
simple tool to extract this quantity.

Waiting for a computational work which will provide us with this information, 
we now try to combine together some known facts in order to get a rough 
estimate of $r^{~}_\R$.  Numerical data on \Cblu{unstable} 
strings in 3+1 dimensional $SU(3)$ \cite{sd,ba} and $SU(2)$ \cite{cp} 
gauge theories seem to support Casimir scaling \cite{aop}, which tells us that
the static potential between sources in the representation $\R$ is proportional 
to that of fundamental sources according to
\eq
V_\R(r)\simeq \frac{{\cal C}_\R}{{\cal C}_f}\,V_f(r)~~,
\label{casimir}
\en
where ${\cal C}_\R$ is the quadratic Casimir operator of the representation 
$\R$.  This implies 
\eq
\sigma_\R^{~} \simeq \frac{{\cal C}_\R}{{\cal C}_f}\,\sigma~,  
~~\mu_\R^{~} \simeq\frac{{\cal C}_\R}{{\cal C}_f} \, \mu~,
\label{cas}
\en
where $\sigma$ and $\mu$ refer to the fundamental representation.
Since the additive UV divergence of $\mu^{~}_\R$ is cancelled by that of  
 $\mu'_\R$ in the difference $\Delta\mu$, it is quite natural to expect 
 $\mu'_\R$ also  proportional to ${\cal C}_\R$. However, only the 
unstable $\R-$strings have a non-vanishing  $\mu'$, hence we assume
\eq
\mu'_\R\simeq\frac{{\cal C}_\R}{{\cal C}_{adj}}\,\,\mu'_{adj}~,
\label{mumu}
\en
where, as noted above, $\mu'_{adj}$ is the mass of the lowest glue-lump.
Thus, combining Eq.s(\ref{scale}),(\ref{cas}) and (\ref{mumu}) leads to
\eq
r_\R^{~}\simeq \,\frac{{\cal C}_\R}
{{\cal C}_\R-\sigma^{~}_k/{\sigma}}\;r_{adj}^{~}~,
\label{esca}
\en
hence  $r^{~}_\R> r_{adj}$ for any \Cblu{unstable} string coupled to 
a colour source with non-vanishing $N-$ality. For instance, in the $SU(2)$ 
case, the $j=\frac32$ is unstable against  decaying into the fundamental 
string with $j=\frac12$, thus $r^{~}_{\frac32}\simeq\frac54 r_{adj}$. 
 
Notice that, unlike the unstable strings, the  
\Cblu{stable} $k-$strings tension 
$\sigma_k/\sigma$ does not obey Eq.(\ref{cas}).  
Numerical experiments 
\cite{dprv,lt,ltw,dprv2} as well as  consistency of $N\to\infty$  limit
\cite{as} point to a sizable
violation of the Casimir scaling and some works \cite{dprv,dprv2} find 
good agreement with the so called sine law discovered in some supersymmetric 
gauge models\cite{ds,hsz,hk}.
\par 
Our rough estimate of the decay lengths of unstable strings were derived under
a number of (strong) assumptions. A better definition, based on the 
multichannel method, will be described in the next section. 
\section{Mixed Wilson loops}
\label{mixed}
To warm up, let us consider the construction of a $SU(N)$ glue-lump operator 
which creates a glue-lump at the point $\vec{x}$ and annihilates it at the 
point $\vec{y}$.
\begin{figure}
\centering
\includegraphics[width=0.95\textwidth]{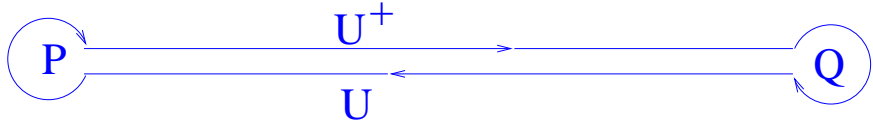} 
\caption{Generating operator of the gluelump, as described in 
Eq.(\ref{glu}).} 
\label{Figure:3}
\end{figure}

The starting point is the gauge-invariant operator depicted in Fig.
\ref{Figure:3}
\eq
P(\vec{x})^i_j\;\;U^\dagger(\vec{x},\vec{y})^j_k\;\; 
Q(\vec{y})^k_l\;\; U(\vec{x},\vec{y})^l_i~~,
\label{glu}
\en
where $P(\vec{x})$ and  $Q(\vec{y})$ are the source and the sink of the 
gluelump. For simplicity, the points $\vec{x}$ and $\vec{y}$ 
are chosen on the same coordinate axis. In a 3D cubic lattice, for instance, 
a good overlap with the 
lowest $0^+$ state is obtained by 
choosing $P$ and $Q$ as the `clover leaves' \cite{mc} formed by the four 
plaquettes orthogonal to  $\vec{y}-\vec{x}$.  $U(\vec{x},\vec{y})$ 
is a shorthand  notation for the parallel transporter formed by the  product 
of the link variables in the fundamental 
representation along the straight path connecting the sites 
$\vec{x}$ and $\vec{y}$. Let us focus on the indices $i$ and $j$ of the 
operators $U$ and $U^\dagger$. They belong to the reducible representation
$\{N\}\otimes\{\bar{N}\}=\{N^2-1\}+\{1\}$. In order to project out the singlet 
and allow propagating the adjoint representation only, we 
perform the substitution
\eq
 {U^\dagger}^j_k \,U^l_i\to   {U^\dagger}^j_k\, U^l_i-\frac1N\,\delta^j_i \,
 {U^\dagger}^m_k\, U^l_m= {U^\dagger}^j_k\, U^l_i-\frac1N\,\delta^j_i\;\delta^l_k~.
\en
Inserting this projection in (\ref{glu}) yields
\eq
G(\vec{x},\vec{y})=\tr(PUQU^\dagger)-\frac1N\,\tr P\;\tr Q~.
\label{gluelump}
\en
In the case of $SU(2)$ we can use the trace identity
\eq
\tr A\,\tr B=\tr(A\,B)+\tr(A\,B^{-1})~,
\label{ti}
\en
valid for any pair of $2\times2$ unimodular matrices
\footnote{Actually it can be shown that the most general solution of 
the functional equation  $\phi(A)\,\phi(B)=\phi(A\,B)+\phi(A\,B^{-1})$, 
where $A$ and $B$ are arbitrary elements of an unspecified group $G$ and 
$\phi$ is a class function, is  $G\equiv SL(2,\C)$ and $\phi$ is the character 
of its fundamental representation. A study of this kind of relationships for various groups can be found in\cite{gv}. }, 
with $A=P$ and $B=UQU^\dagger$ and recast 
Eq.(\ref{gluelump}) in the form used by the lattice community
\eq
G(\vec{x},\vec{y})=\um\tr[PU(Q-Q^\dagger)U^\dagger]\equiv
\um\tr(P\sigma^a)\Gamma_{ab}\tr(Q\sigma^b)~,
\en
where the $\sigma^a\equiv\sigma_a$ are the Pauli matrices , $a=1,2,3$, and
\eq
\Gamma_{ab}=\um\tr(\sigma_aU\sigma_bU^\dagger)~.
\en

\begin{figure}
\centering
\includegraphics[width=0.95\textwidth]{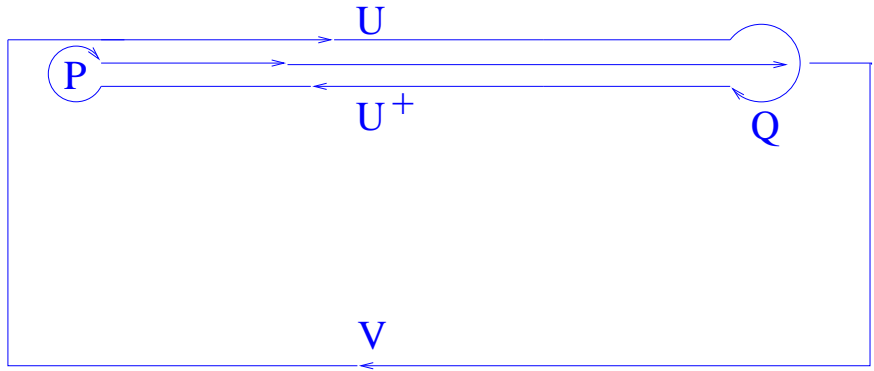} 
\caption{Mixed  Wilson loop. The loop $(PUVUQU^\dagger)$ denotes a static source in the fundamental representation $f$. The  three coincident 
lines $U,U,U^\dagger$ at  the top are drawn separately for clarity. They 
belong to the reducible representation $f\otimes f\otimes \bar{f}$. 
Projection on  a irreducible component is described in the text. } 
\label{Figure:4}
\end{figure}

To construct a \Cblu{mixed Wilson loop} let us start by considering an 
arbitrary closed path $\gamma=u\,v$ made by the composition of two open paths
$u$ and $v$. Let $U$ and $V$ be the group elements associated with 
these two paths respectively. The associated standard Wilson loop is
\eq
W(\gamma)=\tr(UV)~.
\label{wo}
\en  
where the trace is taken , here and in the following, in the fundamental 
representation $f$  of the \Cblu{\sl non-abelian} gauge group $G$. We want to 
transform $W$ in a mixed Wilson loop in which the source along the path $u$ 
carries the quantum numbers of an
higher representation $\R$ belonging to the same $N-$ ality class of $f$.
In analogy with the construction of the gluelump operator, we start 
by considering  the gauge-invariant 
generating operator (see Fig.\ref{Figure:4})
\eq
P^m_i\,U^i_l\,V^l_j\,U^j_k\,Q^k_n\,{U^\dagger}^n_m
\label{genera}
\en  
where $P$ and $Q$ can be taken as the source operators  associated to the 
``clovers'' orthogonal to the path $\gamma$ at the junctions separating $u$
and $v$.    Along the path $u$ now propagates a  source belonging to the 
reducible $f\otimes f\otimes\bar{f} $. We have then to project on 
some irreducible component.

 To make a specific, illustrative example, let us consider the case of 
$SU(3)$, where we have
\eq
\{3\}\otimes \{3\}\otimes\{\bar{3}\}=2\{3\}+\{\bar{6}\}+\{15\}~.
\en
We want to project on the $\{\bar{6}\}$ representation which has the same
triality of $\{3\}$. It may be selected, for instance, by anti-symmetrizing 
the indices $i$ and $j$ and eliminating the traces 
with the index $m$ in the matrix elements $P^m_i\,V^l_j$ . 
Namely, we take the combination

\eq
\frac32(P^m_j\,V^l_i-P^m_i\,V^l_j)+ 
\frac34[(\delta^m_i\,V^l_j -\delta^m_j\,V^l_i) \tr P-(VP)^l_j\,\delta^m_i+
\delta^m_j\,(VP)^l_i]
\label{prj6}
\en
and saturate with $ U^i_l\,U^j_k\,Q^k_n\,{U^\dagger}^n_m$, getting finally
\eq
\begin{split}
W^{~}_{\{3\}\to\{\bar{6}\}}(\gamma)=\frac32\left[\tr(PUQU^\dagger)\,\tr(VU)-
\tr(PUVUQU^\dagger)\right]+\\
\frac34\left[\tr P\, \tr(VUQ)-\tr(VPUQ)+\tr (VPU)\,\tr Q-\tr P\tr(VU)\,\tr Q
\,\right]~.
\label{w36}
\end{split}
\en
The  fraction $u$ of $\gamma$ belonging to $\{\bar{6}\}$ is of course arbitrary. Moving the two junctions along $\gamma$ we can manage as to shrink the 
length $\vert v\vert$ of $v$ to zero, hence $V$  is the identity matrix
$V=\uu$ and the whole loop carries the quantum numbers of $\{\bar{6}\}$.
In this limit the two clovers associated to $P$ and $Q$, which have opposite 
orientations, overlap, hence $P=Q^\dagger$. 
As a check of Eq.(\ref{w36}) we can now 
integrate over $Q$, using the standard orthogonality relations of the 
irreducible characters
\eq
\begin{split}
\int_{Q\in G}dQ\,\Tr_\R(Q\,A)\Tr_{\R'}(Q^\dagger B)=\delta_{\R,\R'}
\,\frac 1{d_\R}\Tr_\R(A^\dagger B)\\
  \int_{Q\in G}dQ\,\Tr_\R(Q\,A\,Q^\dagger B)=
\frac 1{d_\R}\Tr_\R(A^\dagger B)~,
\end{split}
\en
 written for any arbitrary compact group $G$. $\Tr_\R$ is the character,
{\sl i.e.} the trace calculated in the irreducible representation $\R$ and 
$d_\R=\Tr_\R(\uu)$ its dimension.

Putting in  Eq.(\ref{w36}) $V=\uu$, $P=Q^\dagger$ and integrating on $Q$ yields
\eq
\um\left[ (\tr U)^2-\tr(U^2)\right]\tr U^\dagger-\tr U=
\um\left[(\tr U^\dagger)^2+\tr({U^\dagger}^2)\right]=
\Tr_{\{\bar{6}\}}(U)~,
\en
where the $SU(3)$ identity $\um[ (\tr U)^2-\tr(U^2)]=\tr U^\dagger$ and 
its conjugate have been used.

Similarly, in the $SU(2)$ case, the mixed Wilson loop  associated to the 
pair  of representations $j=\um$ and $j=\frac32$ turns out to be
\eq
\begin{split}
W^{~}_{\um\to\frac32}(\gamma)=\tr(PUVUQU^\dagger)+\tr(PUQU^\dagger)\tr(UV)-\\
\frac13\left[\tr(PUQV)+\tr(PUV)\,\tr Q+\tr P\left(\tr(VUQ)+\tr(VU)\,\tr Q
\right)\right]~,
\end{split}
\en
where the normalisation is chosen, like in the $SU(3)$ case, in such a way 
that integration on $Q$ after putting $V=\uu$ and $P=Q^\dagger$ gives
\eq
\int_{Q\in SU(2)}dQ\,W^{~}_{\um\to\frac32}=
\Tr_{\frac32}(U)\equiv\tr(U^2)\,\tr U~.
\en
 \par
It is clear that the above construction can be generalised to any non-abelian 
group and, in particular, to any fully anti-symmetric representation 
of $N-$ality $k$ of $SU(N)$, which can be converted through the emission and 
the reabsorption of a glue-lump to an excited representation $\R$.
It is also clear that one can build up mixed Wilson loops of the 
type drawn in Fig.\ref{Figure:1} $c$ and $d$ that we denote respectively as 
$W^{~}_{k\to\R}(r,t)$ and $W^{~}_{\R\to k\to\R}(r,t)$\footnote{The ``time'' 
variable $t$ is the vertical line in Fig.\ref{Figure:1}.}.
\par The static potential $V_\R(r)$ between the $\R$ sources and the 
decay of the 
associated unstable string into the stable $k-$string can then be extracted 
from measurements of the matrix correlator
\eq
C(r,t)=\left(
\begin{matrix}
\bra W^{~}_{\R}(r,t)\ket &\bra W^{~}_{k\to\R}(r,t)\ket\\
\bra W^{~}_{k\to\R}(r,t)\ket&\bra W^{~}_{\R\to k\to\R}(r,t)\ket
\end{matrix}
\right)~,
\label{cor}
\en   
where $W^{~}_{\R}(r,t)$ is the ordinary Wilson loop  when the whole 
rectangle $(r,t)$ is in the $\R$ representation. This is the 
generalisation of the multichannel method we alluded in the Introduction.
Denoting by $\lambda(r,t)$  the highest eigenvalue of Eq.(\ref{cor})
we have
\eq
V_\R(r)=-\lim_{t\to\infty}\frac1t\log\lambda(r,t)~,
\en
which from a computational point of view is much better than the definition
(\ref{poto}), because the mixed Wilson loops have, by construction, a much 
better overlap with the stable $k-$string state. 
In this way it will possible to evaluate the decay scale $r_\R$. 
One could also try to get a rough estimate of this quantity through 
Eq.(\ref{scale}). Indeed the instability of the $\R-$string leads to conjecture that the vacuum expectation value of 
$W^{~}_{k\to\R}(\gamma)$ should behave asymptotically as 
\eq
\bra W^{~}_{k\to\R}(\gamma)\ket\propto e^{-\mu_\R^{~}\vert v\vert-
\mu_\R'\vert u\vert-\sigma^{~}_k A}~,
\en
where $A$ is the area of the minimal surface encircled by $\gamma$ and 
$\vert v\vert$ and $\vert u\vert$  are the lengths of the paths which carry 
 the quantum numbers of the representations $k$ and $\R$, respectively.
This seems the most effective way to estimate the quantity $\mu'_\R$
and therefore $r^{~}_\R$.
 \section{Conclusion}
\label{conclusion}
In this paper we gained some insight into the physics of $SU(N)$ pure
gauge theory by using in two different ways the standard decomposition 
of the direct product of irreducible representations of the gauge group.

First, we mapped this decomposition into the fusion rules of the effective CFT 
describing the critical behaviour of the finite-temperature deconfinement 
 of those 2+1 dimensional gauge theories which undergo a continuous phase 
transition. 

We worked out two specific examples, $SU(3)$ and $SU(4)$, which 
led us to conjecture the exact critical exponents to be attributed to Polyakov 
lines in some higher representations of the gauge group. 
The resulting functional form of these Polyakov-Polyakov correlators at 
critically suggests generalising it to $T=0$ Wilson loops in higher 
representations for any $SU(N)$. The proposed Ansatz offers a simple, 
general explanation of the difficulty to observe decaying unstable strings 
while measuring Wilson loops only.

The other way we exploited the mentioned decomposition in irreducible 
representations has been used to define a new gauge-invariant operator,
the mixed Wilson loop $W^{~}_{\R\to\s} (\gamma)$ , which describes a 
source belonging to an irreducible representation $\R$ which is converted into 
another representation $\s$ 
along part of its trajectory $\gamma$. Gauge invariance requires $\R$ and 
$\s$ having the same $N-$ality. This new operator could be used as a tool 
to study the decay of the unstable strings, a recurrent theme of this work.

\end{document}